# Transmission Techniques for Relay-Interference Networks

Soheil Mohajer *, Suhas N. Diggavi*, Christina Fragouli *, David Tse †

* School of Computer and Communication Sciences
École Polytechnique Fédéral de Lausanne, Lausanne, Switzerland.
Email: {soheil.mohajer, suhas.diggavi, christina.fragouli}@epfl.ch
† Wireless Foundations, UC Berkeley, Berkeley, California, USA.
Email: dtse@eecs.berkeley.edu

*Abstract*— In this paper we study the relay-interference wireless network, in which relay (helper) nodes are to facilitate competing information flows over a wireless network. We examine this in the context of a deterministic wireless interaction model, which eliminates the channel noise and focuses on the signal interactions. Using this model, we show that almost all the known schemes such as interference suppression, interference alignment and interference separation are necessary for relay-interference networks. In addition, we discover a new interference management technique, which we call interference neutralization, which allows for over-the-air interference removal, without the transmitters having complete access the interfering signals. We show that interference separation, suppression, and neutralization arise in a fundamental manner, since we show complete characterizations for special configurations of the relay-interference network.

*Index Terms*— Interference channel, wireless relay networks, multiple unicast, deterministic channel, interference neutralization.

## I. INTRODUCTION

Information transmission in a shared medium is one of the fundamental problems in wireless communication. In such situation a wireless channel is shared between several sources and receivers, and several information flow are competing for resources. Here, a fundamental question is how to manage interference in a wireless network.

In the multiple access channel problem, introduced by Ahlswede and Liao in early 70's, a single receiver is interested in decoding the messages sent by different transmitters. Several techniques, including multi-user detection, orthogonal source allocation, and taking interference as a part of noise have been devised for this problem.

In a more general setup, not all the source are of interest for all the receivers. The interference channel problem [1] is the very basic example of such situation which has been open for 30 years. The best known achievable region for this problem is due to Han and Kobayashi [1]. Over the past few decades several techniques have been devised for transmission on the interference channels; among them, superposition of information, power allocation, and interference suppression (partly common information) are the most well-known ones. Recently, the capacity region of the interference channel has been characterized for some regimes by building on an approximate characterization (within 1 bit) given for the whole regime in [2]. However, it is not clear whether the known techniques are enough to achieve the capacity when we also have relays in the network facilitating the flow of more than one unicast session.

The deterministic approach, studied by Avestimehr, Diggavi, and Tse [3], [4], simplifies the wireless network interaction model by eliminating the noise. This approach was successfully applied to the relay network in [4], and resulted in insight in terms of transmission techniques. These insights also led to an approximate characterization of the noisy wireless relay network problem [5]. This model is also applied to the interference channel problem in [6], where it is shown that the capacity region of the deterministic interference channel is within constant bit gap of the Gaussian interference channel, and an alternative approximate characterization for the capacity region is provided.

In this paper, we apply the deterministic model to a two-stage interference channel, where the goal is to accommodate multiple unicast flows over the network. The simple layered structure of the networks helps us to focus more on the transmission techniques, rather than synchronization issues, raised in a non-layered network. We have complete characterization for two special cases, called the ZS and the ZZ networks. Investigation of these networks, suggest a new insight about the transmission techniques, which can be applied in any network. It is shown that the interference separation and interference suppression are useful to avoid or remove interference in different regimes. We will also show that using interference alignment is essential for some cases, even with two messages transmitted through the network. The other contribution of this paper is to introduce a new transmission technique, *interference neutralization*, to remove (decrease) the interference in a network.

The paper is organized as follows. Section II states the precise definition of the problem, and introduces the notations. Before stating the main results, we review the known techniques and explain the new techniques that we will use later in Section III. We will present our main results, the exact characterization of the ZS and ZZ networks in Sections IV and V. Finally, we will conclude and discuss about future extensions in Section VI.

This work was supported in part by the Swiss National Science Foundation through NCCR-MICS under grant number 51NF40-111400 and the FNSF award number 200021-103836/1.

## II. NOTATIONS AND PROBLEM STATEMENT

*Wireless interaction model:* In this standard model [7], transmitted signals get attenuated by (complex) gains to which independent (Gaussian) receiver noise is added. More formally, the received signal $y_i$ at node $i \in \mathcal{V}$ at time $t$ is given by,

$$y_i(t) = \sum_{j \in \mathcal{N}_i} h_{ij} x_j(t) + z_i(t), \quad (1)$$

where $h_{ij}$ is the complex channel gain between node $j$ and $i$, $x_j$ is the signal transmitted by node $j$, and $\mathcal{N}_i$ are the set of nodes that have non-zero channel gains to $i$. We assume that the average transmit power constraints for all nodes is 1 and the additive receiver Gaussian noise is of unit variance. We use the terminology *Gaussian wireless network* when the signal interaction model is governed by (1).

*Deterministic interaction model:* In [4], a simpler deterministic model which captures the essence of wireless interaction was developed. The advantage of this model is its simplicity, which gives insight to strategies for the noisy wireless network model in (1). We will utilize this model to develop techniques for the relay-interference network. Our main results are developed for this deterministic model. The deterministic model of [4] simplifies the wireless interaction model in (1) by eliminating the noise and discretizing the channel gains through a binary expansion of $q$ bits. Therefore, the received signal $Y_i$ which is a binary vector of size $q$ is modeled as

$$Y_i(t) = \sum_{j \in \mathcal{N}_i} M_{ij} X_j(t), \quad (2)$$

where $M_{ij}$ is a $q \times q$ binary matrix representing the (discretized) channel transformation between nodes $j$ and $i$ and $X_j$ is the (discretized) transmitted signal. All operations in (2) are done over the binary field, $\mathbb{F}_2$. We use the terminology *deterministic wireless network* when the signal interaction model is governed by (2). Shift matrix is a special matrix representation for a Gaussian fading channel. This matrix captures the attenuation effect of the signal caused by the channel gain by performing a shift on the binary representation of the input, $x_j$, and ignoring the bits below the average noise level. More precisely, this model assigns a matrix $J^{q-n_{ij}}$ to the Gaussian gain $h_{ij}$, where

$$J = \begin{pmatrix} 0 & 0 & 0 & \cdots & 0 \\ 1 & 0 & 0 & \cdots & 0 \\ 0 & 1 & 0 & \cdots & 0 \\ \vdots & \ddots & \ddots & \ddots & \ddots \\ 0 & \cdots & 0 & 1 & 0 \end{pmatrix}_{q \times q}, \quad (3)$$

is the shift matrix, and $n_{ij} = \lceil \frac{1}{2} \log |h_{ij}|^2 \rceil$, for real channel gains.

An illustration of this deterministic model is given in Fig. 1 for the broadcast and multiple access networks. Fig. 1(a) shows a deterministic model of the broadcast channel, where the channel from the transmitter to Receiver 1 is stronger than that to Receiver 2. This is represented by the deterministic model developed in [4] with 4 most significant bits (MSB) of the transmitted signal captured by $D_1$ and only 2 MSBs of the transmitted signal captured by $D_2$. The deterministic model of the multiple access channel shown in Fig. 1(b) adds one more ingredient, which is how the bits from two transmitting nodes interact at a receiver. In Fig. 1(b) the channel from $S_1$ to $D$ is stronger than that of $S_2$. Therefore, the interaction is between the 2 MSBs of the message sent by $S_2$ with the lower 2 significant bits of the message sent by $S_1$, and the interaction is modeled with an addition over the binary field (*i.e.*, xor). This interaction captures the dynamic range of the signal interactions. It was shown in [4], that this model *approximately*[1] captures the wireless interaction model of (1) for the broadcast and multiple access channels. For general networks the deterministic model yields insights which, when translated to the noisy wireless network, lead one to develop cooperative strategies for the model in (1), which are (provably) approximately[2] optimal [5].

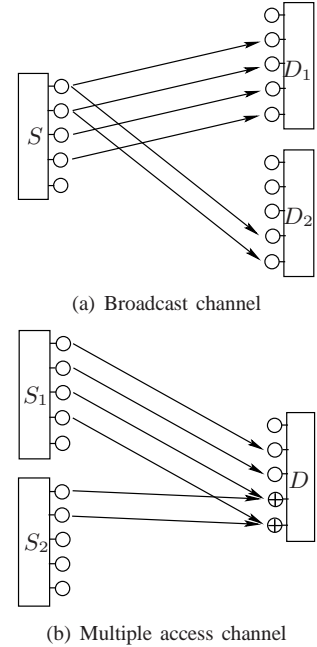

(a) Broadcast channel

(b) Multiple access channel

Fig. 1. The linear deterministic model for a Gaussian broadcast channel (BC) is shown in (a) and for a Gaussian multiple access channel (MAC) is shown in (b).

### A. Multi-unicast Deterministic Network

Our goal is to characterize the capacity region of a network with two unicast sessions under the deterministic shift model. A simple example of such network is a two stage layered interference network shown in Fig. 2, which we call it the XX network. There are two transmitters $S_1$ and $S_2$ which encode their messages $W_1$ and $W_2$ of rates $r_1$ and $r_2$, respectively, and broadcast the obtained vectors to the relay nodes, $R_1$ and $R_2$.

---
[1]The approximation is in the sense that the capacity region of the deterministic model is within 1 bit of the capacity region of the Gaussian counterparts.

[2]It has been shown for single unicast there is an *approximate* max-flow, min-cut result where the difference is within a constant number of bits, which depends on the topology of the network, but not the values of the channel gains [5].

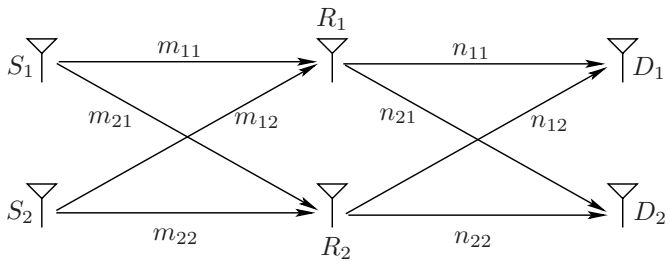

Fig. 2. Transmission model

We assume that the messages are encoded using a code of length $n$, and denote the transmitted vectors in time block $T$ (time instants $\{(T-1)n+1, (T-1)n+2\ldots, Tn\}$) by $X_1^n(T)$ and $X_2^n(T)$. The relay node $R_1$ ($R_2$) receives a signal $Y_1^{'n}(T)$ ($Y_2^{'n}(T)$) which is deterministic function of the vectors sent by the transmitters, as in a multiple access channel. Here, $m_{ij}$ denotes the channel gain from $S_j$ to $R_i$, for $i, j \in \{1, 2\}$. The transmission model of the first stage of the network could be summarized by

$$Y_1' = Y_{11}' + Y_{12}' = M_{11}X_1 + M_{12}X_2, \quad (4)$$
$$Y_2' = Y_{21}' + Y_{22}' = M_{21}X_1 + M_{22}X_2, \quad (5)$$

where $M_{ij} = J^{n-m_{ij}}$ is a power of the shift matrix, and $Y_{ij}'$ is the message received by relay node $R_i$ from the source $S_j$ when there is no interference.

The relay nodes wait until the end of the time block and apply a proper function on the set of vectors received in time block $T$, and broadcast the resulting vectors in time block $T+1$. We denote by $X_1^{'n}(T+1)$ and $X_2^{'n}(T+1)$ the vectors transmitted by relays in this time block. In time block $T+1$, destination nodes $D_1$ and $D_2$ receive interfered signal $Y_1^n(T+1)$ and $Y_2^n(T+1)$ which depend on both $X_1^{'n}(T+1)$ and $X_2^{'n}(T+1)$. The channel gain from relay node $R_j$ to destination node $D_i$ is denoted by $n_{ij}$, for $i, j \in \{1, 2\}$. More precisely,

$$Y_1 = Y_{11} + Y_{12} = N_{11}X_1' + N_{12}X_2', \quad (6)$$
$$Y_2 = Y_{21} + Y_{22} = N_{21}X_1' + N_{22}X_2', \quad (7)$$

and $N_{ij} = J^{n-n_{ij}}$.

Getting all the $n$ vectors in time block $T+1$, the destination nodes decode the messages sent by the source nodes in time block $T$. Destination node $D_i$ is only interested in decoding message $W_i$. We may drop the time block indicators ($T$ or $T+1$) whenever it is clear from the context, and does not cause confusion.

A rate pair $(r_1, r_2)$ is called admissible if there exist a scheme for a large enough $n$, where $D_1$ and $D_2$ can decode $W_1$ and $W_2$, respectively. It is worth mentioning that this network acts like a two stage cascaded interference network. However, the important difference here is that, unlike in the interference network, the messages sent by the relays at the second phase of transmission are not independent. This fact affects on the capacity region of the network.

In this paper, instead of studying the admissible rate region of the general network, the main focus is on two specific realization of the network, namely, the capacity regions of the ZS and the ZZ networks. Our main goal is to illustrate the transmission techniques utilized in order to achieve such capacity regions and we illustrate some of the more general networks.

### III. Transmission Techniques

In this section we illustrate some examples, each of which benefits from one of the techniques we have mentioned in the last section.

#### A. Interference Separation

Consider the network shown in Fig. 3. It is easy to see that the sum-rate of this network is upperbounded by

$$r_1 + r_2 \leq 3, \quad (8)$$

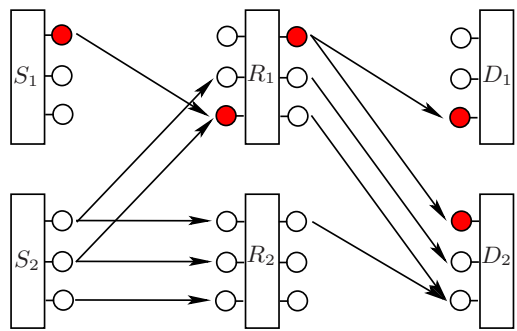

Fig. 3. Interference separation; $(r_1, r_2) = (1, 2)$ is achievable.

by studying the cut-set which separates the destination nodes from the rest of the network.

Assume we wish to transmit at rate pair $(r_1, r_2) = (1, 2)$ from the source nodes to the destination nodes. It turns out that this can be done only using an opportunistic encoding which avoid interference. Since $D_1$ receives only one bit from $R_1$, this bit should be the clear data about $W_1$. Hence, $R_1$ should have received the message from $S_1$, without interference. Therefore, the message $W_2$ should be encoded such that it does not cause interference at $R_1$. More precisely, in order to to communicate at this rate, the transmitters should encode their messages as

$$X_1 = \begin{pmatrix} x_1(1) \\ 0 \\ 0 \end{pmatrix}, \quad X_2 = \begin{pmatrix} x_2(1) \\ 0 \\ x_2(2) \end{pmatrix}, \quad (9)$$

and the relay node have to perform proper linear operations on their received signal before broadcasting them.

$$X_1' = \begin{pmatrix} x_1(1) \\ x_2(1) \\ 0 \end{pmatrix}, \quad X_2' = \begin{pmatrix} x_2(2) \\ 0 \\ 0 \end{pmatrix}. \quad (10)$$

It is clear this encoding scheme makes the interference separable from the signal at the nodes $R_1$ and $D_2$. It can be seen that this is necessary as well for this example.

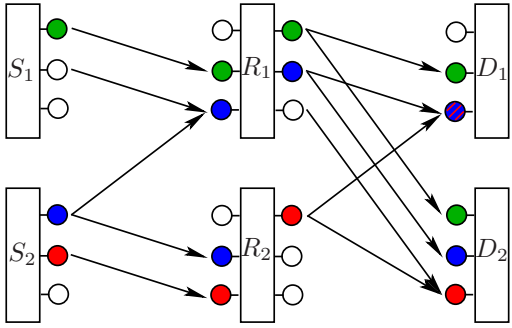

Fig. 4. Interference alignment; $(r_1, r_2) = (1, 2)$ is achievable.

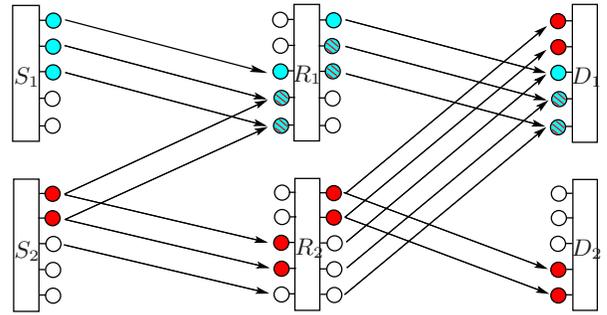

Fig. 5. Interference suppression: $(r_1, r_2) = (3, 2)$ is achievable.

### B. Interference Alignment

Interference alignment is a transmission technique to put all the interference into a small dimension. The intuition behind this technique is the fact that whatever scheme one uses to deal with interference, it occupies a certain number of degrees of freedom of the network. Therefore, one could expect to minimize such loss by aligning the footprints of all interferences so that the total interference is the smallest. Such phenomenon has been observed in a X channel [8] wherein two transmitters attempt to communicate to two receivers, over an interference channel. Each transmitter has two messages, each to be decoded at one receiver. It has been shown that in order to achieve the capacity of this network, it is necessary to align the two signals carrying information about the irrelevant interfering messages at the receivers. In [8], this phenomenon has been observed for the case where there are more than two messages have to be transmitted in the network. However, in this subsection, we show through an example that interference alignment might be an essential strategy even with two messages in our relay-interference network.

Consider the network shown in Fig. 4. We wish to communicate at rate pair $(r_1, r_2) = (1, 2)$. Since there is only one link from $R_2$ to $D_2$, it is clear that the relay node $R_1$ should help them by sending information bits about $X_2$. Therefore, the destination node $D_1$ receives two interfering signal (from $R_1$ and $R_2$) which describe $X_2$. Hence, it would be able to resolve $X_1$ if and only if the occupied sub-node by these two interference coincide. More precisely, encoding the messages as

$$X_1 = \begin{pmatrix} x_1(1) \\ 0 \\ 0 \end{pmatrix}, \qquad X_2 = \begin{pmatrix} x_2(1) \\ x_2(2) \\ 0 \end{pmatrix}, \qquad (11)$$

the received signal at the destination nodes using the shown transmission strategy would be

$$Y_1 = \begin{pmatrix} 0 \\ x_1(1) \\ x_2(1) + x_2(2) \end{pmatrix}, \qquad Y_2 = \begin{pmatrix} x_1(1) \\ x_2(1) \\ x_2(2) \end{pmatrix}. \qquad (12)$$

This shows that the interfering bits $x_2(1)$ and $x_2(2)$ are aligned at the destination node $D_1$.

### C. Interference Suppression

Depending on the parameters of the network, there are cases in which neither interference separation nor interference alignment are optimal to achieve a transmission rate pair. Namely, there is no way to avoid interference at the receivers. However, it might be possible to receive a clean copy of the interference beside the copy who interfered the signal. Therefore, one could use the clean copy to remove the interference. This is exactly the situation can be be observed in the ZZ network shown in Fig. 5.

In this network the signal observed at $R_1$ is interfered by $S_2$, and there is no way for $R_1$ to decode $W_1$ when the transmission rate is $(r_1, r_2) = (3, 2)$. In order to achieve this rate pair, the decoder $D_1$ has to first (partially) decode $W_2$ using the message received from $R_2$, and then use this message to remove the interference from the signal received from $R_1$. This is the only strategy we can use to decode $W_1$ at $D_1$. Note that here there are two interfering paths from $S_2$ to $D_1$. However, the second path (through $R_2$) helps the decoder to remove the interference caused by the first path (through $R_1$).

### D. Interference Neutralization

This technique can be used in networks which contain more than one disjoint path from $S_i$ to $D_j$ for $i \neq j$, where $D_j$ is not interested in decoding the message sent by the source node $S_i$, and therefore it receives the interference through more than one link. The proposed technique is to tune these interfering signals such that they neutralize each other at the destination node. In words, the interfering signal should be received at the same power level and with different sign such that the effective interference, obtained by adding them, occupies a smaller number of degrees of freedom. This technique is new and has not been considered in the literature up to best of our knowledge.

Fig. 6 shows a network in which interference neutralization is essential to achieve the desired rate pair $(r_1, r_2) = (2, 3)$. Here $D_1$ has only two degrees of freedom, and receives information bits from both $R_1$ and $R_2$ over these sub-nodes. However, notice that there are two disjoint paths $(S_2, R_1, D_1)$ and $(S_2, R_2, D_1)$, which connect $S_2$ to $D_1$. Using a proper mapping (permutation) at the relay nodes, one can make the interference neutralized at the destination node $D_1$, and provide two non-interfered links from $S_1$ to $D_2$. Note that this permutation does not effect the admissible rate of the other unicast from $S_2$ to $D_2$, the cost we pay, is to re-permute the received bits at $D_2$.

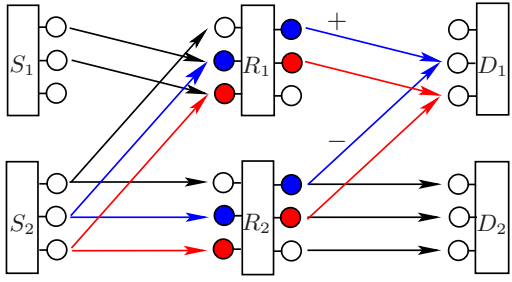

Fig. 6. Interference neutralization; $(r_1, r_2) = (2, 3)$ is achievable.

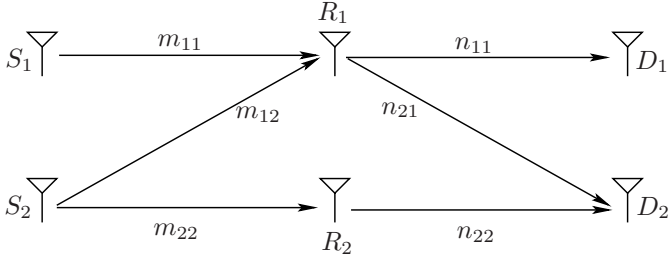

Fig. 7. The ZS network

## IV. THE ZS NETWORK

In this section, we restrict our attention on a specific network by assuming zero gain for two of the cross links. This assumption leads us to the network shown in Fig. 7. In the following we obtain the admissible rate region of this network under the deterministic model and restrict our analysis to the shifting matrices.

The following theorem gives a complete characterization of the capacity region of the ZS network.

*Theorem 1:* The admissible rate region of the ZS network is the set of all rate pairs $(r_1, r_2)$ which satisfy

$$r_1 \leq m_{11}, \qquad \text{(ZS-1)}$$
$$r_1 \leq n_{11}, \qquad \text{(ZS-2)}$$
$$r_2 \leq \max(m_{12}, m_{22}), \qquad \text{(ZS-3)}$$
$$r_2 \leq \max(n_{21}, n_{22}), \qquad \text{(ZS-4)}$$
$$r_2 \leq m_{12} + n_{22}, \qquad \text{(ZS-5)}$$
$$r_2 \leq m_{22} + n_{21}, \qquad \text{(ZS-6)}$$
$$r_1 + r_2 \leq \max(m_{11}, m_{12}) + n_{22}, \qquad \text{(ZS-7)}$$
$$r_1 + r_2 \leq m_{22} + \max(n_{11}, n_{21}), \qquad \text{(ZS-8)}$$
$$r_1 + r_2 \leq \max(m_{11}, m_{12}) + (m_{22} - m_{12})^+, \qquad \text{(ZS-9)}$$
$$r_1 + r_2 \leq \max(n_{21}, n_{22}) + (n_{11} - n_{21})^+ \qquad \text{(ZS-10)}$$

In the following subsections we briefly state the outline of the proof of the optimality as well as the achievability of this rate region.

### A. Necessity: The Proof of the Converse

In this subsection we show that any achievable rate pair $(r_1, r_2)$ satisfies (ZS-1)-(ZS-10). All the inequalities in the theorem above, are essentially obtained using the maximum-flow min-cut theorem. Clearly, the inequalities (ZS-1) and (ZS-2) bound the flow of information from $S_1$ and to $D_1$, respectively. The inequality given by (ZS-3) is simply the bound for broadcasting data from $S_2$. Similarly, (ZS-4) is the multiple access sum-rate bound for the destination node $D_2$.

In order to prove (ZS-5), we consider the cut which partitions the network into $\Omega_s = \{S_2, R_2\}$ and $\Omega_d = \{S_1, R_1, D_1, D_2\}$. We have

$$nr_2 \leq I(Y_1'^n(T), Y_2^n(T+1); X_2'^n(T+1), X_2^n(T))$$
$$= I(Y_1'^n; X_2'^n, X_2^n) + I(Y_2^n; X_2'^n, X_2^n | Y_1'^n)$$
$$= I(Y_1'^n; X_2^n) + H(Y_2^n | Y_1'^n) \qquad (13)$$
$$\leq I(Y_1'^n; X_2^n | X_1^n) + H(Y_2^n | X_1'^n) \qquad (14)$$
$$= H(Y_1'^n | X_1^n) + H(Y_2^n | X_1'^n)$$
$$= H(Y_{12}'^n) + H(Y_{21}^n)$$
$$\leq \mathsf{rank}(M_{12}) + \mathsf{rank}(N_{22})$$
$$= m_{12} + n_{22}, \qquad (15)$$

where in (13) we used the fact that $X_2'^n(T+1)$ is a function of $X_2^n(T)$, and (14) holds since $X_2^n(T)$ is independent of $X_1^n(T)$, and $X_1'(T+1)$ is a function of $Y_1'(T)$.

The inequality (ZS-6) can be similarly obtained by bounding the information flow through the cut $\Omega_s = \{S_1, S_2, R_1, D_1\}$ and $\Omega_d = \{R_2, D_2\}$.

The inequality (ZS-7) captures the information flow through the cut which partitions the network into $\Omega_s = \{S_1, S_2, R_2\}$ and $\Omega_d = \{R_1, D_1, D_2\}$. The maximum flow of information through this cut can be evaluated as

$$n(r_1 + r_2) \leq I(Y_1'^n(T), Y_2^n(T+1);$$
$$\qquad X_1^n(T), X_2^n(T), X_2'^n(T+1))$$
$$= I(Y_1'^n; X_1^n, X_2^n, X_2'^n)$$
$$\quad + I(Y_2^n; X_1^n, X_2^n, X_2'^n | Y_1'^n)$$
$$\leq H(Y_1'^n) + H(Y_2^n | Y_1'^n)$$
$$\leq H(Y_1'^n) + H(Y_2^n | X_1'^n)$$
$$\leq H(Y_1'^n) + H(Y_{22}^n)$$
$$\leq n\mathsf{rank}\begin{bmatrix} M_{11} & M_{12} \end{bmatrix} + n\mathsf{rank}(N_{22})$$
$$= n\max(m_{11}, m_{12}) + nn_{22} \qquad (16)$$

Similarly, we can prove (ZS-8) by bunding the information flow through the cut $\Omega_s = \{S_1, S_2, R_1\}$ and $\Omega_d = \{R_2, D_1, D_2\}$.

It remains to show prove the upperbounds (ZS-9) and (ZS-10). Consider the cut $\Omega_s = \{S_1, S_2\}$ and $\Omega_d = \{R_1, R_2, D_1, D_2\}$. The flow of information through this cut

can be upper bounded as

$$n(r_1 + r_2) \leq I(Y_1'^n(T), Y_2'^n(T); X_1^n(T), X_2^n(T))$$
$$= I(Y_1'^n; X_1^n, X_2^n) + I(Y_2'^n; X_1^n, X_2^n | Y_1'^n)$$
$$\leq I(Y_1'^n; X_1^n, X_2^n) + H(Y_2'^n | Y_1'^n)$$
$$\quad - H(Y_2'^n | X_1^n, X_2^n, Y_1'^n)$$
$$\leq n \max(m_{11}, m_{12}) + H(Y_2' | Y_1') \quad (17)$$

Note that $D_1$ receives information only through $R_1$. Therefore since $D_1$ is able to decode $W_1$, so $R_1$ is. Hence, using Fano's inequality we can write

$$H(Y_2'^n | Y_1'^n) \leq H(Y_2'^n | Y_1'^n, W_1) + n\varepsilon$$
$$= H(Y_2'^n | Y_{11}'^n, Y_{12}'^n) + n\varepsilon$$
$$\leq H(Y_2'^n | Y_{12}'^n)$$
$$= n(m_{22} - m_{12})^+. \quad (18)$$

The proof of inequality (ZS-10) follows a similar argument by bounding the information flow through the cut $\Omega_s = \{S_1, S_2, R_1, R_2\}$ and $\Omega_d = \{D_1, D_2\}$.

$$n(r_1 + r_2) \leq I(Y_1^n(T+1), Y_2^n(T+1);$$
$$\qquad X_1'^n(T+1); X_2'^n(T+1))$$
$$= H(Y_1^n, Y_2^n)$$
$$= H(Y_2^n) + H(Y_1^n | Y_2^n)$$
$$\leq n \max(n_{21}, n_{22}) + H(Y_1^n | Y_2^n). \quad (19)$$

Now, we can use the facts that having $Y_2^n(T+1)$, $D_2$ is assumed to be able to decode $W_2$. Also it is clear that $X'^n(T+1)$ is a deterministic function of $W_2$, and therefore that of $Y_2^n(T+1)$.

$$H(Y_1^n | Y_2^n) \leq H(Y_1^n | Y_2^n, W_2) + n\varepsilon$$
$$\leq H(Y_1^n | Y_2^n, Y_{22}^n) + n\varepsilon$$
$$\leq H(Y_1^n | Y_{21}^n)$$
$$\leq (n_{11} - n_{21})^+ \quad (20)$$

*B. Achievability*

Here we only give the outline of the transmission scheme that can be used to achieve the rates given in Theorem 1. This scheme works by decomposition of the network into two *isolated* components, namely $\mathcal{N}_1$ and $\mathcal{N}_2$, where each pair of source/destination uses one component for communication. Here by two isolated network we mean two networks which are completely disjoint and and no message can be transmitted from one to another. The fact that the components are isolated, guarantees that the signals do not cause interference.

Such decomposition is based on the desired transmission rate pair. We denote by $S_1(\mathcal{N}_i)$ the set of subnodes at $S_1$ which are included in the network component, $\mathcal{N}_i$, and use similar notation for partitioning the other subnodes of the network. In order to communicate at rate $(r_1, r_2)$, which satisfies (ZS-1)-(ZS-10), we form the components of the network as follows. $S_1(\mathcal{N}_1)$ includes the top $(m_{11} - m_{12})^+$ as well as the lowest $(r_1 - (m_{11} - m_{12})^+)^+$ subnodes of $S_1$. This component also includes any receiver node from $R_1$ which is connected to a node in $S_1(\mathcal{N}_1)$, and any node in $D_2$ which is connected to a node in $R_1(\mathcal{N}_1)$. Clearly the receiver nodes in $R_2$ which are connected to a node in $D_2(\mathcal{N}_1)$ are also included in $\mathcal{N}_1$.

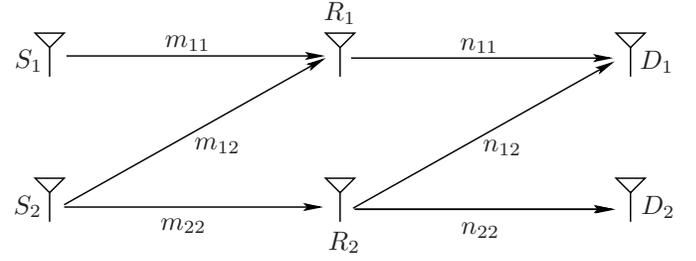

Fig. 8. The ZZ network.

Similarly, the transmitter part of $R_1$ in the first component includes the top $(n_{11} - n_{21})^+$ as well as the lowest $(r_1 - (n_{11} - n_{21})^+)^+$ nodes of the transmitter side of $R_1$. In consequence, all the corresponding subnodes in $D_1$, $R_2$ and $D_2$ are also included in $\mathcal{N}_1$. The second component of the network is formed by all the remaining subnodes in the network.

It is clear the this decomposition is isolated by its construction, and the $S_1/D_1$ can use the first component to communicate at $r_1$ as a line network. One can also characterize the second network $\mathcal{N}_2$, as network with two relays and show that $S_2/D_2$ can communicate over this network at rate $r_2$. We deligate the proof details of this part to the journal version because of lack of space.

## V. THE ZZ NETWORK

In this section we consider another special case of the XX network. Here, we assume that the cross-links from $S_1$ to $R_2$, and also the cross link from $R_1$ to $D_2$ have zero gain. Therefore, the remaining network would be two Z network which are cascaded as shown in Fig. 8.

In the following we will obtain the admissible rate region of a ZZ network. The capacity of a single Z channel under deterministic shift model has been computed in [6]. However, it turns out that the rate region of a ZZ network could be a strict superset of the rate region of the single Z network. It is not surprising, since the messages broadcasted by the relay nodes in the secand stage of the network are not independent, while in a Z network they are.

The key observation of this enlargement of the capacity region is the interference neutralization. In fact, in a single Z network the cross link acts as an interference for one of the receivers. In a ZZ network as shown in Fig. 8, the signal received at $D_1$ is effected by two interference (through links $S_2$ to $R_1$ and $R_2$ to $D_1$) which carry information about the same message $X_2$. Therefore, it is possible that to encode them such the total effective interference be weaker than the original one. In other words, the interference caused by the link $S_2$ to $R_1$ can be (partially) neutralized by the other cross link in the second stage of the network. More precisely, the signals sent by the relays can be amplified properly such that they have opposite effect on the effective interference at $D_1$, and therefore partially neutralize each other.

In the following theorem, we characterize the admissible rate region of a ZZ network.

*Theorem 2:* The capacity of region of a deterministic ZZ network shown in Fig. 8 is given be the set of all $(r_1, r_2)$ which satisfy

$$r_1 \leq m_{11}, \tag{ZZ-1}$$
$$r_2 \leq m_{22}, \tag{ZZ-2}$$
$$r_1 \leq n_{11}, \tag{ZZ-3}$$
$$r_2 \leq n_{22}, \tag{ZZ-4}$$
$$r_1 + r_2 \leq \max(m_{11}, m_{12}) + (m_{22} - m_{12})^+ + n_{12}, \tag{ZZ-5}$$
$$r_1 + r_2 \leq \max(n_{11}, n_{12}) + (n_{22} - n_{12})^+ + m_{12}\}. \tag{ZZ-6}$$

*A. Necessity: The Proof of the Converse*

In this section we briefly state the proof of the optimality of the rate region introduced in Theorem 2. The first inequality (ZZ-1) is simply obtained by the maximum information flow through the cut $\Omega_s = \{S_1\}$ and $\Omega_d = \{S_2, R_1, R_2, D_1, D_2\}$.

$$nr_1 \leq I(X_1^n; Y_1'^n | Y_{12}'^n)$$
$$= H(Y'n_1 | Y'n_{12})$$
$$\leq H(Y_{11}'^n)$$
$$\leq n\text{rank}(M_{11}) = nm_{11}. \tag{21}$$

Similarly $r_2$ can be upper bounded by the information flow through the cut set $\Omega_s = \{S_1, S_2, R_1, R_2, D_1\}$ and $\Omega_d = \{D_2\}$ as in (ZZ-4).

Note that (ZZ-2) is tighter than the cut-set bound. In fact, $D_2$ receives information only from $R_2$. So whatever $D_2$ can decode is also decodable by $R_2$. This can be seen through the Markov chain $X_2^n(T) \leftrightarrow Y_2'^n(T) \leftrightarrow X_2'^n(T+1) \leftrightarrow Y_2^n(T+1)$.

$$nr_2 \leq H(W_2) \leq I(W_2; Y_2^n) + H(W_2|Y_2^n)$$
$$\leq I(X_2^n; Y_2^n) + n\varepsilon_n \tag{22}$$
$$\leq I(X_2^n; Y_2'^n) + n\varepsilon_n$$
$$= H(Y'n_2) + n\varepsilon$$
$$\leq n\text{rank}(M_{22}) = nm_{22} + n\varepsilon, \tag{23}$$

where (22) follows from the Fano's inequality.

We can also upper bound $r_1$ as in (ZZ-3) by noting the fact that the only path in the network which connects $S_1$ to $D_1$ passes through $R_1$. Therefore, the Markov chain $Y_2 \leftrightarrow (X_1', X_2') \leftrightarrow (X_1, X_2')$ holds.

$$nr_1 \leq I(Y_1^n; X_1) \leq I(Y_1^n; X_1 | X_2'^n) \tag{24}$$
$$= H(Y_1^n | X_2'^n) - H(Y_1^n | X_1, X_2')$$
$$= H(Y_1^n | X_2'^n) - H(Y_1^n | X_1', X_2')$$
$$\leq nH(Y_1 | X_2'), \tag{25}$$

where (24) holds since the message sent by $R_2$ only depends on $X_2^n$ and is independent of $X_1^n$. Finally, we have

$$r_1 \leq H(Y_1 | X_2') \leq \text{rank}(N_{11}) = n_{11}. \tag{26}$$

The proof of the sum-rate bounds are more technical. For each inequality we start with the information flow through a cut-set, and then we use a key observation in evaluating the cut-set value.

Consider the cut-set $\Omega_s = \{S_1, S_2\}$ and $\Omega_d = \{R_1, R_2, D_1, D_2\}$. In order to prove (ZZ-5), we provide the information sent by $R_2$ to $D_1$ for $R_1$ as side information. In such condition, the information $R_1$ has about $W_1$ is stronger than the information $D_1$ has, and therefore $R_1$ can decode $W_1$. By removing the interference from $S_1$, $R_1$ can also get partial information about $W_2$. More precisely, we can write

$$n(r_1 + r_2) \leq I(X_1^n(T), X_2^n(T); Y_1'^n(T), Y_2'^n(T))$$
$$= H(Y_1'^n, Y_2'^n)$$
$$\leq H(Y_1'^n, Y_2'^n, Y_{12}^n(T+1))$$
$$= H(Y_1'^n, Y_{12}^n) + H(Y_2'^n | Y_1'^n, Y_{12}^n)$$
$$\leq H(Y_1'^n) + H(Y_{12}^n) + H(Y_2'^n | Y_1'^n, Y_{12}^n)$$
$$\leq \max(m_{11}, m_{12}) + n_{12} + H(Y_2'^n | Y_1'^n, Y_{12}^n) \tag{27}$$

Note that

$$H(Y_{12}'^n | Y_1'^n, Y_{12}^n) = H(Y_1'^n - Y_{11}'^n | Y_1'^n, Y_{12}^n)$$
$$\leq H(Y_1'^n, Y_{11}'^n | Y_1'^n, Y_{12}^n)$$
$$= H(Y_{11}'^n | Y_1'^n, Y_{12}^n)$$
$$\leq H(W_1 | Y_1'^n, Y_{12}^n)$$
$$\leq H(W_1 | Y_{11}^n, Y_{12}^n)$$
$$\leq H(W_1 | Y_{11}^n + Y_{12}^n))$$
$$= H(W_1 | Y_1^n) \leq n\varepsilon_n \tag{28}$$

where $\varepsilon_n \to 0$ as $n$ grows. Note that (28) follows from the Fano's inequality, and the fact that $D_1$ can decode the message sent by $S_1$. Hence,

$$H(Y_2' | Y_1', Y_{12}) \leq H(Y_2', Y_{12}' | Y_1', Y_{12})$$
$$= H(Y_2' | Y_{12}', Y_1', Y_{12}) + H(Y_{12}' | Y_1', Y_{12})$$
$$\leq H(Y_2' | Y_{12}') + \varepsilon_n$$
$$\leq (m_{22} - m_{12})^+ + \varepsilon_n. \tag{29}$$

Continuing from 27 we have

$$r_1 + r_2 \leq H(Y_1') + H(Y_{12}) + H(Y_2' | Y_1', Y_{12})$$
$$\leq \max(m_{11}, m_{12}) + n_{12} + (m_{22} - m_{12})^+. \tag{30}$$

The last inequality (ZZ-6), intuitively means that the number of neutralized sublinks at $D_1$ cannot exceed the minimum of $m_{12}$ and $n_{12}$. We can similarly prove it by considering the information flow through the cut $\Omega_s = \{S_1, S_2, R_1, R_2\}$ and $\Omega_d = \{D_1, D_2\}$, and providing $D_1$ by the information sent by $S_2$ to $R_1$ as side information.

$$n(r_1 + r_2) \leq I(Y_1^n(T+1), Y_2^n(T+1);$$
$$X_1'^n(T+1); X_2'^n(T+1))$$
$$= H(Y_1^n, Y_2^n)$$
$$\leq H(Y_1^n, Y_2^n, Y_{12}'^n(T))$$
$$\leq H(Y_1^n) + H(Y_{12}'^n) + H(Y_2^n | Y_1^n, Y_{12}'^n)$$

Similar to proof of (ZZ-5), we use the following bounding technique.

$$\begin{aligned} H(Y_{12}^n|Y_1^n, Y_{12}'^n) &= H(Y_{12}^n - Y_{11}^n|Y_1^n, Y_{12}'^n) \\ &\leq H(Y_1^n, Y_{11}^n|Y_1^n, Y_{12}'^n) \\ &= H(Y_{11}^n|Y_1^n, Y_{12}'^n) \\ &\leq H(Y_1'^n|Y_1^n, Y_{12}'^n) \\ &= H(Y_{11}'^n + Y_{12}'^n|Y_1^n, Y_{12}'^n) \\ &= H(Y_{11}'^n|Y_1^n, Y_{12}'^n) \\ &\leq H(Y_{11}'^n|Y_1^n) \\ &\leq H(W_1|Y_1^n) \leq n\varepsilon_n, \end{aligned} \quad (31)$$

where we have used the Fano's inequality in (31). Therefore,

$$\begin{aligned} H(Y_2^n|Y_1^n, Y_{12}'^n) &\leq H(Y_2^n, Y_{12}^n|Y_1^n, Y_{12}'^n) \\ &= H(Y_2^n|Y_{12}^n, Y_1^n, Y_{12}'^n) + H(Y_{12}^n|Y_1^n, Y_{12}'^n) \\ &\leq H(Y_2^n|Y_{12}^n) + \varepsilon_n \\ &\leq (n_{22} - n_{12})^+ + \varepsilon_n. \end{aligned} \quad (32)$$

Therefore, we have

$$r_1 + r_2 \leq \max(n_{11}, n_{12}) + m_{12} + (n_{22} - n_{12})^+. \quad (33)$$

*B. Achievability*

In this subsection we briefly state the transmission technique that can be used to achieve the capacity region of a ZZ network. The keypoint here is to use the capability of interference neutralization and interference suppression suggested by the two disjoint paths from $S_2$ to $D_1$ to overcome the interference effects.

Our scheme is based on a network decomposing. We say that four subnodes $a \in S_1$, $b \in R_1$, $c \in S_2$, and $d \in R_2$ form a *full Z path* if $c$ broadcasts to $b$ and $d$, and $b$ receives info mation from $a$ and $c$ as in a multiple access channel. It can be shown that the number of full Z paths in first layer of the network is given by

$$\delta_{SR} = \min(m_{11}, m_{12}, m_{22}, (m_{11} + m_{22} - m_{12})^+). \quad (34)$$

Similarly we form the full Z paths in the second layer of the network. The number of such paths would be

$$\delta_{RD} = \min(n_{11}, n_{12}, n_{22}, (n_{11} + n_{22} - n_{12})^+). \quad (35)$$

Therefore, we have $\delta = \min(\delta_{SR}, \delta_{RD})$ *full ZZ pairs*, in the network.

The decomposing of the network, forms two components, one containing the maximum number of full ZZ pairs, and the other one which is the rest of the network. It is easy to show that these two networks are isolated and do not cause interference to each other. Moreover, the first component, the full ZZ pairs, form two paths from $S_1$ to $D_1$ and from $S_2$ to $D_2$, with $\delta$ degrees of freedom. The interference in this component is *natural* neutralization by the structure of the network, and therefore, it forms $\delta$ clean links for each pair of source/destination.

Depending on the parameters of the network, $\mathcal{N}_2$, the second component can be either a cascade of a Z channel with two parallel links, a network with only one transmitter (receiver), or a ZZ network wherein no more full ZZ pair exists. It is easy to study this component and obtain an optimal transmission strategy in the two former cases. However, it requires more technical details to investigate the second component if the remaining is a ZZ network. It can be shown that in this situation we neet to apply either interference neutralization or interference suppression to convey maximum amount of information through the network. Using such strategies we can show that in any of the mentioned case, any rate pair satisfying Theorem 2 is achievable.

## VI. CONCLUSION

In this paper we examined the relay-interference network, which is a natural combination of the relay network along with the interference channel. This can also be thought of as the multiple-unicast problem in the context of wireless networks. The broadcast nature of wireless communication makes signal interactions more complicated, leading to a challenging problem. We make progress on this question by studying it using the deterministic model introduced in [4]. We show that besides the known interference management techniques such as interference suppression, alignment and separation, we also need a new technique we term interference neutralization. The characterization for two-stage ZZ and ZS networks demonstrate that this new technique arises in a fundamental manner. In ongoing work, we have made progress on approximate characterizations of the noisy (Gaussian) version of the relay-interference networks. We hope to completely answer this and questions related to arbitrary configurations for the two-unicast problem in a future work.